# Magneto-transport properties in the thin films of charge ordered materials

A.Venimadhav[a], M.S Hegde[a], R.Rawat* and I.Das*

a)Solid state and structural chemistry unit, Indian institute of science, Bangalore-12, India

*IUC center for DAE facility, Indore, India

## Abstract

Thin films of $Pr_{0.7}Ca_{0.3}MnO_3$, $Nd_{0.5}Ca_{0.5}MnO_3$ and $Nd_{0.5}Sr_{0.5}MnO_3$ have been fabricated by pulse laser deposition. Magnetic and electrical transport properties of these films were compared with their bulk solids. All the films grow in (101) direction on $LaAlO_3$(100). Electrical transport measurements carried out with and without magnetic field has shown $d\rho/dT<0$ in these thin films. Application of magnetic field has shown dilution of the insulating behavior. The magnetization studies of $Nd_{0.5}Sr_{0.5}MnO_3$ exhibited ferromagnetic behavior at 260K and showed antiferromagnetic feature below 130K. This is close to the bulk magnetic behavior of $Nd_{0.5}Sr_{0.5}MnO_3$. $Pr_{0.7}Ca_{0.3}MnO_3$ showed ferromagnetic transition below 130K and becomes antiferromagnetic at 90K. $Nd_{0.5}Ca_{0.5}MnO_3$ showed paramagnetic behavior down to low temperatures. Deviation in the properties of the thin films from the bulk solids is attributed to the growth of the films in more symmetric structure.

mshegde@sscu.iisc.ernet.in

madhav@sscu.iisc.ernet.in

# I. Introduction

There has been an intense research activity on the hole doped manganite perovskite oxides for the last few years because of their interesting collosal magnetoresistance (CMR) property. Manganite perovskites $LnMnO_3$ (Ln = La, Nd, Pr and Sm) are antiferromagnetic (AFM) insulators. On hole doping, $Ln_{1-x}A_xMnO_3$ (A = Ca, Sr, Pb etc) for $0.2 < x < 0.5$, they become ferromagnetic metal (FMM) and show CMR effect. The change in the transport property is coupled with the structural transition in these materials. On increasing the doping concentration, crystal structure changes from distorted orthorhombic to more symmetric rombohedral[1]. There have been extensive experimental and theoretical studies to understand this origin of the anomalous magneto transport phenomenon and qualitative explanation is given by double exchange (DE) mechanism. At higher doping levels ( $x > 0.5$), the ground state becomes again an AFM insulator. Another intriguing phase namely the charge ordered (CO) state has been found to exist in insulating $Ln_{1/2}A_{1/2}MnO_3$. The CO state is characterized by the real space ordering of $Mn^{3+}/Mn^{4+}$ in the mixed valent manganite. In the case of Ln = Pr, A = Ca, charge ordering is observed for $0.3 < x < 0.5$[2]. The CO state is expected to become stable when the coulombic repulsion between the carriers dominates over the kinetic energy of the carriers. The FMM state can be reentered from the CO state by applying external magnetic fields which is termed as the melting of the CO state. The melting of the CO state is associated with a huge MR of thousand folds.

$Pr_{0.5}Ca_{0.5}MnO_3$ and $Nd_{0.5}Ca_{0.5}MnO_3$ crystallize in the orthorhombic structure with a space group Pnma. $Nd_{0.5}Sr_{0.5}MnO_3$ crystallizes in Imma. In the bulk, NSMO is a FMM with $T_c$ of 250 K, which transforms to CO AFM state below 140K[3]. This CO state can be melted by the application of 8T magnetic field. PCMO has a paramagnetic ground state at room temperature. On cooling it goes to CO state at 240 K and becomes AFM below 150 K. The CO state is associated by a large jump in the resistivity. High magnetic field is needed to melt the CO state. In case of $Nd_{0.5}Ca_{0.5}MnO_3$ the CO insulating state is followed by the PMI state at 260 K. There are a few reports on the thin film study of $Pr_{0.5}Ca_{0.5}MnO_3$ , $Nd_{0.5}Sr_{0.5}MnO_3$ and $Pr_{0.5}Sr_{0.5}MnO_3$ CO materials[4-6]. The melting of the CO insulating state to metallic state is not observed in $Pr_{0.5}Ca_{0.5}MnO_3$ thin films on $LaAlO_3$[4,7]. Since the direct application of these materials is likely to be in the thin film technology, there is need for the study of these materials in thin films. In this paper we report the structural, magnetic, electrical transport and MR property of $Nd_{0.5}Sr_{0.5}MnO_3$, $Pr_{0.7}Ca_{0.3}MnO_3$ and $Nd_{0.5}Ca_{0.5}MnO_3$ thin films prepared by pulse laser deposition.

## II. Experiment

Stoichiometric pellets of the $Nd_{0.5}Sr_{0.5}MnO_3$ (NSMO) $Pr_{0.7}Ca_{0.3}MnO_3$ (PCMO), and $Nd_{0.5}Ca_{0.5}MnO_3$ (NCMO) and have been synthesized by solid state route. Polycrystalline samples were characterized by X-ray diffraction, electrical transport and magnetic measurements in order to confirm their phases. Thin film of PCMO, NCMO and NSMO were prepared by pulse laser deposition on $LaAlO_3$ (100). All the films were grown at a substrate temperature of 780 $^0$C. The deposition is carried out in flowing oxygen atmosphere of 300mT. The films were annealed for one hour at 750 $^0$C in oxygen pressure of 500Torr. The thickness of the grown films were ~3000 Å. The structure of the films were characterized by X- ray diffraction using Seimens D5000 power diffractometer. DC Resistivity measurements were done by four probe method. Magnetic measurements were performed using Faraday force balance. Manetoresistance measurements were done up to a magnetic field of 8T. Magnetic field is applied parallel to the substrate surface and the direction of the current flow.

## III. Result and Discussion

Fig 1 shows the XRD $\theta$-$2\theta$ pattern of the NSMO, PCMO and NCMO and films. X ray diffraction pattern of NSMO film gave an out of plane parameter of 3.86 Å. This value is close to the bulk $d_{101}$. Thin film of NSMO reported on LAO is also shown to grow along (101) direction[5]. The out of plane parameter for the PCMO film is calculated to be 3.85 Å. A similar value is reported on LAO by Perillier et al. for PCMO film[4]. This value is close to $d_{101}$ of the bulk solids. Hence the PCMO film grows along the (101) direction perpendicular to LAO (100). The NCMO film also gave an out of plane parameter of 3.85 Å. This is expected as both PCMO and NCMO are structurally similar and also their lattice parameters are close. From Fig.1 we conclude that the films were highly oriented and grows along (101) direction.

Resistivity measurements of these samples with and without magnetic field were shown in Fig 2. The resistivity of the NSMO showed $d\rho/dT<0$. No charge ordering behavior is seen in the resistivity. The R vs T behavior was different and could not be fitted to either variable range hopping or thermal activation. Both PCMO and NCMO showed $d\rho/dT < 0$ activation process as the conduction mechanism. Inset of Fig. 2 shows the fitting of the zero field data of PCMO film to the thermal activation process. We have not observed charge ordering in PCMO and NCMO films which is expected to occur at 240 K for PCMO and a slight shoulder at 260 K for NCMO respectively in the bulk samples. From the resistivity data it is clear that the conductivity of NSMO film is more than PCMO and NCMO films. This is also true in the bulk solids as NSMO falls in the higher average A site ionic radii of the CO materials[8]. It is also clear that CO is not observed in these films. Thus the insulating behavior could be attributed to the localization of the conducting carriers. The effect of magnetic field on the resistivity of the thin films is shown in the Fig. 2. PCMO and NSMO films showed a considerable decrease of their resistivity by the application of an external magnetic field. The application of the magnetic field reduces the localization of the carriers. Though with the application of 8T magnetic field on these films no metallic behavior was observed. The NCMO film exhibited considerable decrease in resistivity with 8T magnetic field. The field dependent resistivity of PCMO and NCMO followed the thermal activation process. Such an activation process was not observed in the NSMO films.

Magnetization measurements on these films were shown in Fig. 3. The measurements were done with a magnetic field of 500 Oe up to 25K. ZFC and FC of NSMO film showed ferromagnetic transition at 265 K. In the ZFC antiferromagnetic transition below 140 K was observed. At 160 K there is a different slope in the FM regime showing up in the magnetization curve. In the bulk solid at this temperature AFM and CO establishes simultaneously. The FM and AFM ordering temperatures in the film were closely matching with the bulk solid value of NSMO. PCMO sample showed a clear curie temperature at 140 K and AFM behavior starting at 90K. AFM behavior also persists in the FC data. The NCMO film showed paramagnetic behavior down to 25K. Magnetoresistance (MR) measurements were done on these films up to a magnetic field of 8T is shown in Fig. 4. The MR of NSMO sample shown in Fig. 4(a) was exhibiting a linear dependence on H even up to low temperatures. This kind of behavior was seen in CMR materials near the transition temperature where the coexistence of PM and FM phases present. This could be attributed to the presence of both PM and FM phases in the film. At low temperature NSMO film showed a hysterisis of MR as shown in the inset of Fig 4.(a). Hysterisis in MR at low temperatures is seen in the bulk solids of the CO materials. The MR of PCMO was shown in Fig 4(b). MR measurements were carried out above and below the curie temperature. The MR plot in the paramagnetic regime has showed $H^2$ dependence. In the FM region at 110K the observed MR was large and exhibited exponential decay with field as observed in CMR materials below $T_c$. The MR behavior with field for NCMO (Fig 4 (c)) showed $H^2$ dependence up to measurable temperatures as expected for PM regime.

It is evident from the above set of results that the bulk properties are not reproduced in the thin films. In the charge ordered materials a small variation in the composition can destroy the charge ordering. For example with a slight deviation in the composition of $Nd_{0.5}Sr_{0.5}MnO_3$ to $Nd_{0.52}Sr_{0.48}MnO_3$[9] will result in FMM state. However the growth parameters in PLD can be controlled to obtain highly stoichiometric films[10]. Accordingly the properties (such as $T_c$, $T_{IM}$ and MR values) of $La_{0.7}Ca_{0.3}MnO_3$ films grown here under identical conditions has the same values as that of the bulk solids. Similarly Na and Cd doped $LaMnO_3$ films[11] reproduced the same bulk solids values. Therefore we believe that the stoichiometry in our films are same as the bulk solids. Hence the observed transport properties which are different from the bulk solids are carefully examined below.

Lattice parameters of the bulk $Nd_{0.5}Sr_{0.5}MnO_3$ is, a = 5.43 Å, b = 7.63 Å and c = 5.47 Å. The out of plane parameter of the film observed here is 3.86 Å, which is slightly more than the d $_{101}$ of the solid ((a/√2 +b/√2)/2). However b/2 value of 3.81 Å is close to the $a_c$ = 3.788 Å of LAO (100). Therefore NSMO film grows in (101) direction of over the (100) of LAO. Hence the NSMO film is slightly elongated in a and c, and compressed along b direction. A similar conclusion of lattice effects on the PCMO films grown on LaAlO$_3$ (100) substrate is arrived by detailed analysis of TEM and XRD studies[4]. The out of plane parameter of the PCMO and NCMO are also larger than the $d_{101}$ of the respective bulk values. Therefore we propose that the films of all the three materials are strained as described above. Due to the lattice strain the growth prefers in a higher symmetric tetragonal structure to the orthorhombic. This implies that the substrate strain brings down the orthorhombic lattice distortion suppressing the charge ordering. Though PCMO and NCMO show activation behavior in the conductivity, their activation energies are small implying more conductivity in the films. Reduced distortion in the film could also lead to such a behavior. Of the three films, NSMO is more conducting than the other two. The application of the magnetic field has an effect on the electrical resistivity in the films which implies the dilution of the insulating state. Another possibility that can lead to the observed behavior could be the electronic phase separation in the material[12]. Recently there are a few reports discussing this complex scenario in the CO manganites[13]. The coexistence of FMM clusters in the PMI matrix can give a magnetization as seen for NSMO film. By applying external magnetic field the percolative path across the FMM clusters can reduce the electrical resistivity.

In conclusion we have made thin films of PCMO, NCMO and NSMO and studied their magneto transport property. Films grow in (101) direction LAO (100). All films showed d$\rho$/dT < 0. Resistivity of PCMO and NCMO films were fitted to polaronic hopping mechanism. NSMO film showed ferromagnetic behavior from 260 K and exhibiting AFM feature below 130 K. PCMO film showed ferromagnetic below 130 K and becomes AFM at 90 K. NSMO remain PM down to low temperatures. The observed behavior is attributed to the strain effects and the formation of FM and PM clusters in the films.

**ACKNOWLEDGEMENTS**

We thank Department of Science and Technology, Government of India for the financial support.

Fig. 1 X ray diffraction $\theta$ - $2\theta$ pattern for thin films of (a) $Nd_{0.5}Sr_{0.5}MnO_3$, (b) $Pr_{0.7}Ca_{0.3}MnO_3$ (c) $Nd_{0.5}Ca_{0.5}MnO_3$

Fig. 2 Resistivity vs Temperature plot of $Nd_{0.5}Sr_{0.5}MnO_3$ $Pr_{0.7}Ca_{0.3}MnO_3$ and $Nd_{0.5}Ca_{0.5}MnO_3$ thin films with and without magnetic field. Inset shows the thermal activation behavior fit for $Pr_{0.7}Ca_{0.3}MnO_3$ film.

Fig. 3 Magnetization vs temperature plot for $Pr_{0.7}Ca_{0.3}MnO_3$, $Nd_{0.5}Ca_{0.5}MnO_3$ and $Nd_{0.5}Sr_{0.5}MnO_3$ thin films

Fig. 4 Magnetoresistance vs applied magnetic field behavior of (a) $Nd_{0.5}Sr_{0.5}MnO_3$, (b) $Pr_{0.7}Ca_{0.3}MnO_3$ (c) $Nd_{0.5}Ca_{0.5}MnO_3$ films

Fig. 1

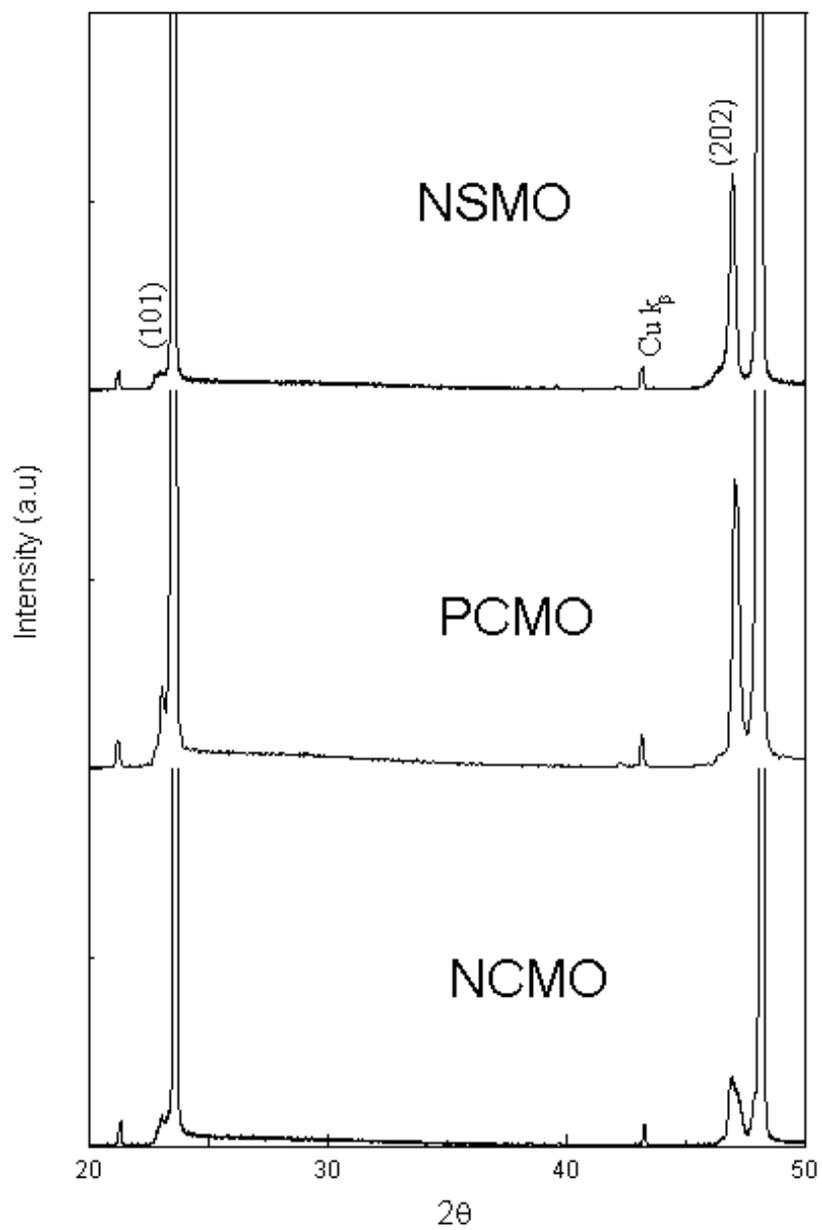

Fig2

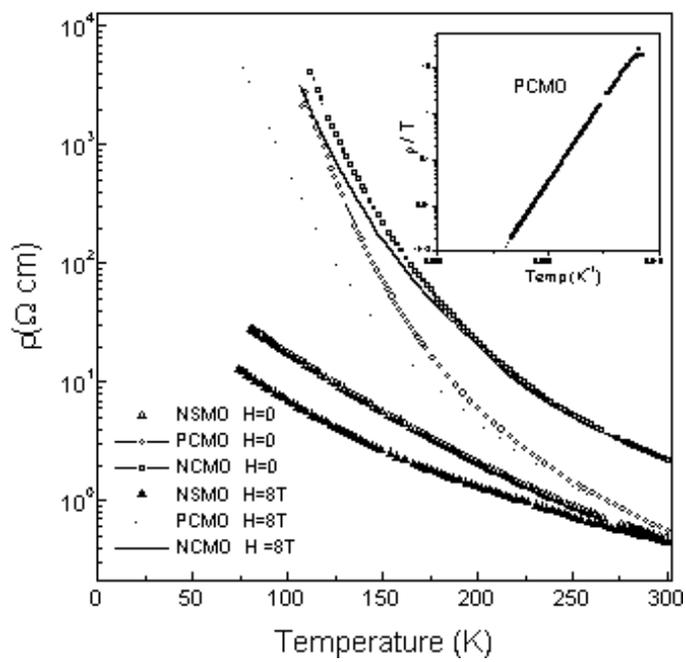

fig 3

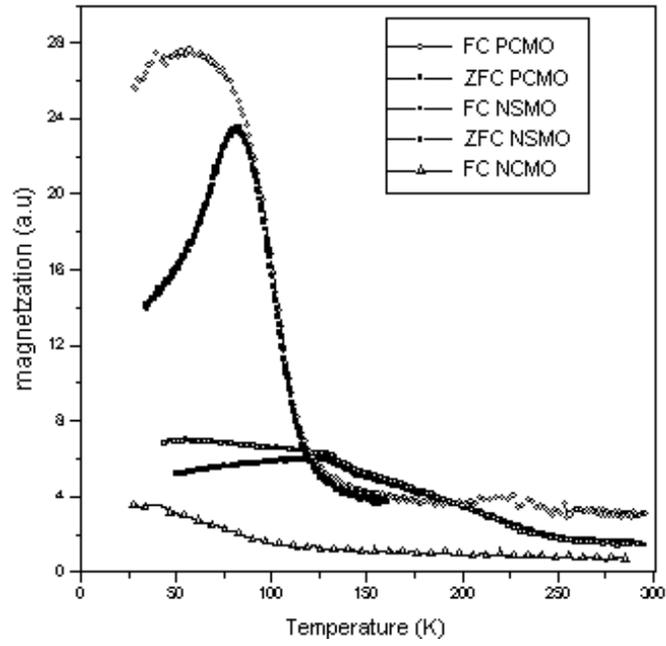

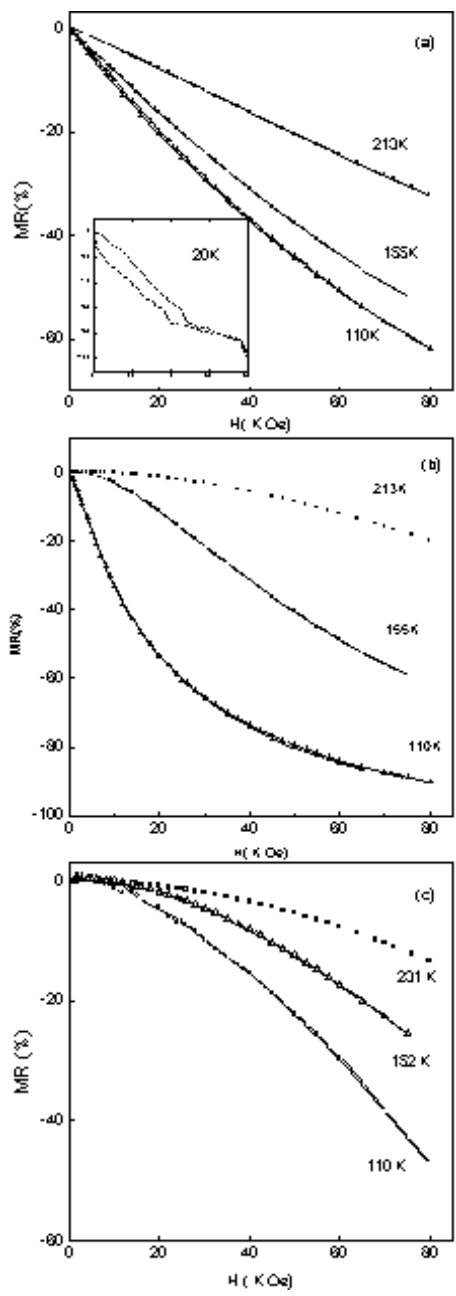

Fig 4